# Implementation of Tripartite Estimands Using Adherence Causal Estimators Under the Causal Inference Framework

Yongming Qu
Department of Biometrics, Eli Lilly and Company, Indianapolis, IN 46285, USA
Email: qu_yongming@lilly.com

Junxiang Luo
Biostatistics and Programming, Sanofi, 55 Corporate Dr, Bridgewater, NJ 08807, USA
Email: junxiang.luo@sanofi.com

Stephen J. Ruberg
Analytix Thinking, LCC, 11121 Bentgrass Court, Indianapolis, IN 46236, USA
Email: AnalytixThinking@gmail.com

May 28, 2020

**Abstract**

Intercurrent events (ICEs) and missing values are inevitable in clinical trials of any size and duration, making it difficult to assess the treatment effect for all patients in randomized clinical trials. Defining the appropriate estimand that is relevant to the clinical research question is the first step in analyzing data. The tripartite estimands, which evaluate the treatment differences in the proportion of patients with ICEs due to adverse events, the proportion of patients with ICEs due to lack of efficacy, and the primary efficacy outcome for those who can adhere to study treatment under the causal inference framework, are of interest to many stakeholders in understanding the totality of treatment effects. In this manuscript, we discuss the details of how to estimate tripartite estimands based on a causal inference framework and how to interpret tripartite estimates through a phase 3 clinical study evaluating a basal insulin treatment for patients with type 1 diabetes.

# 1. Introduction

From the passage of the landmark Kefauver Harris Amendment of the United States Federal Food, Drug and Cosmetic Act [1] to the present day, increasing rigor has been required in the statistical design, analysis, and interpretation of adequate and well-controlled trials. ICH-E9: Statistical Principles for

Clinical Trials [2] provides guidance on statistical principles in study design and data analysis. More recently, the issuance of the National Research Council report on The Prevention and Treatment of Missing Data in Clinical Trials [3] explicitly highlights the notion of the estimand. As noted in that treatise, "Estimation of the primary (causal) estimand, with an appropriate estimate of uncertainty, is the main goal of a clinical trial". ICH E9 (R1), a recent addendum to ICH-E9, discusses estimands in greater detail, especially the role of intercurrent events (ICEs) in defining estimands [4]. ICH E9 (R1) states, "Intercurrent events are events occurring after treatment initiation that affect either the interpretation or the existence of the measurements associated with the clinical question of interest". ICH E9 (R1) provides 5 types of strategies to handle ICEs in defining the estimand: treatment policy strategies, hypothetical strategies, composite variable strategies, while on treatment strategies, and principal stratum strategies. All strategies except the principal stratum strategies intend to define an estimand for all patients. In addition, there are limited discussions on the estimands related to safety, especially for the estimands related to ICEs themselves.

As the assessment of any new treatment is always a balance between treatment effects related to both benefit and risk, Akacha, Bretz and Ruberg [5] have proposed the tripartite approach – a set of three estimands that they argue are meaningful not only to patients, prescribers, and payers, but also sponsors and regulators. Their tripartite estimands are:

1. Probability of patients discontinuing study treatment due to adverse events (AE)
2. Probability of patients discontinuing study treatment due to lack of efficacy (LoE)
3. The treatment effect in patients who can adhere to treatment

For Estimand 3, the adherence status is a post-randomization variable that may depend on the assigned treatment and, consequently, the adherence sets for the 2 treatment groups may not be comparable. The causal inference framework is required to define the estimand and construct a corresponding estimator. Qu et al. propose a theoretical framework for adherence causal estimators (ACEs) based on the causal-inference framework for the estimands of Estimand 3 [6], but they do not provide details on the implementation of tripartite estimands.

In this article, we will clarify the tripartite estimands using the terminology in ICH E9 (R1). Specifically, we define the Estimands 1 and 2 for the proportion of patients with their first ICEs due to AE and LoE (not just treatment discontinuation) and define Estimand 3 using the principal stratum strategies in the causal inference framework [7]. We will then provide details on how to implement the tripartite estimands in clinical trials through retrospective analyses of a phase 3 study. The article is organized in the following way. Section 2 will discuss the statistical methods including the refinement of the tripartite estimands and review the statistical method for the estimation of Estimand 3 as presented by Qu et al. [6].

Section 3 will cover information relevant to the clinical case study, including the data and specific details for our implementation of the tripartite analysis, and will present the analysis results. Finally, Section 4 will discuss our learnings, additional considerations, and potential for further research. For the purposes of this manuscript, we will use the term "treatment" to cover drugs, biologics, or any well-defined intervention and will focus on the clinical trial or development program for such treatments, most notably confirmatory trials that usually constitute phase 3 of clinical development.

## 2. Methods

For a treatment that has a prolonged effect on a primary efficacy measure or is a disease-modifying agent, even data collected after ICEs retain some measure of the randomized treatment effect even with the discontinuation of that randomized treatment or in the presence of rescue medication and can be used to assess the causal effect of the treatment (e.g., disease-modifying antirheumatic drugs or DMARDs). For a treatment that has a more immediate offset of effect on a primary response variable (e.g., a treatment that temporarily ameliorates symptoms), data collected after the cessation of treatment may be additionally confounded by the ICEs, and thus may not be appropriate to be directly used in assessing the causal effect of that treatment. In this article, we address the latter situation.

To define the tripartite estimands, we first introduce the notations of variables of interest and potential outcomes. For intercurrent events, we let $D_{\max}$ be the designed duration of treatment for the study, $D_{AE}$ denote the time to the occurrence of an ICE due to AE, $D_{LoE}$ denote the time to the occurrence of an ICE due to LoE, and $D_A$ denote the time to the occurrence of an ICE due to administrative reasons. Additionally, let $Y$ denote the primary outcome, $Z$ be the intermediate outcome on which the intercurrent events and/or $Y$ may depend, $T$ be the treatment indicator (0 for the reference treatment and 1 for the experimental treatment), and $A$ be the adherence indicator (0 for not being adherent [i.e., with ICEs] and 1 for being adherent [i.e., without ICEs]). Then, the adherence indicator is given by

$$A = I(D_{AE} > D_{\max}) \cdot I(D_{LoE} > D_{\max}) \cdot I(D_A > D_{\max}).$$

For any variable, the potential outcome under a treatment $T$ is denoted by "$(T)$" following the variable. For example, $Y(1)$ denotes the potential outcome under the experimental treatment.

Incorporating the intercurrent events terminology in ICH E9 (R1) and the principal stratum strategies [7-14] for the estimand, a set of tripartite estimands is defined as:

1. For all patients, the treatment difference in the probability of patients with the first ICE due to adverse events (AE)

2. For all patients, the treatment difference in the probability of patients with the first ICE due to lack of efficacy (LoE)
3. For patients who are in a principal stratum defined by treatment adherence, the difference in the primary efficacy outcome.

The analysis variable ($I_{AE}$) for the first estimand is the occurrence (yes or no) of the first ICE due to AE, and the analysis variable ($I_{LoE}$) for the second estimand is the occurrence (yes or no) of the first ICE due to LoE. Then, the variables $I_{AE}$ and $I_{LoE}$ can be defined as

$$I_{AE} = \begin{cases} 1 & D_{AE} \leq D_{\max} \text{ and } D_{AE} \leq D_{LoE} \text{ and } D_{AE} \leq D_A \\ 0 & \text{Otherwise} \end{cases} \tag{1}$$

and

$$I_{LoE} = \begin{cases} 1 & D_{LoE} \leq D_{\max} \text{ and } D_{LoE} \leq D_{AE} \text{ and } D_{LoE} \leq D_A \\ 0 & \text{Otherwise} \end{cases}. \tag{2}$$

Based on the definitions in (1) and (2), the sets $\{I_{AE} = 1\}$ and $\{I_{LoE} = 1\}$ are not exclusive. A patient may discontinue a treatment due to both AE and LoE when $D_{AE} = D_{LoE}$. The first and second estimands can be mathematically expressed as

$$e_1 = E[I_{AE}(1) - I_{AE}(0)] \tag{3}$$

and

$$e_2 = E[I_{LoE}(1) - I_{LoE}(0)], \tag{4}$$

respectively. Based on the definitions in (1) and (2), $I_{AE}$ and $I_{LoE}$ are observed for every patient. Therefore, estimation of the first 2 estimands is very straightforward. Note those with ICEs due to administrative reasons (e.g., moving to a new house, inconvenience due to job changes) are independent of treatment and the primary outcome given baseline covariates. Therefore, the baseline characteristics for patients with ICEs due to administrative reasons should be balanced between treatment, and an estimand for patients who discontinue due to administrative reasons does not seem relevant and is not pursued in the tripartite approach. Note ICEs due to administrative reasons may depend on baseline covariates, but it will not create imbalance between treatment groups.

For Estimand 3, the adherence principal stratum can be defined as adhering to one of the randomized treatments or both randomized treatments, depending on the clinical population of interest.

A general form of the third estimand can be described as

$$e_3 = E\{Y(1) - Y(0)|S\}, \tag{5}$$

where S can be any principal stratum based on adherence. The estimation of the treatment effect for such a principal stratum was first proposed by Frangakis and Rubin [8], and there have been subsequent overviews and discussions of estimation for principal stratum [10, 12, 15]. Generally, there are 4 approaches in estimating $e_3$ in (5):

- Methods based on monotonicity assumption [9, 13, 16-19]
- Methods that estimate the boundary, assume some sensitivity parameter(s), or use the Bayesian method to pose restriction on the sensitivity parameters [13, 14, 20, 21]
- Methods based on estimating the principal score (the probability belonging to a principal stratum) based on baseline covariates [18, 22, 23]
- Methods based on the estimation of potential outcome and/or principal score via baseline covariates and potential post-baseline intermediate measurements [7, 11]

All the above methods require relatively strong assumptions. The monotonicity assumption, such as $A(1) \geq A(0)$ or $A(1) \leq A(0)$, is unreasonable because it assumes a deterministic inequality between 2 random variables. The adherence status $A(t)$ is a random variable that cannot be observed when $t$ is not equal to the assigned treatment $T$ in parallel studies. For a given study population, if we repeat the clinical trial (random treatment assignment and subsequent follow-up) many times, the set of adherers for each treatment will likely be different for each realization of the repeated clinical trial. This is fundamentally different from all randomized patients, which are fixed once the enrollment completes. For convenience, we will still use "population" or "stratum" to describe such a random set of adherent patients, but in all statistical derivations, we treat $A$ as a random variable.

Methods for estimating the boundary of the estimand on a principal stratum and methods depending on sensitivity parameters can only be used for sensitivity analysis because they either cannot provide a consistent estimator, or the sensitivity parameter cannot be objectively estimated. Methods of estimation using a principal score based on baseline covariates are also unlikely to be sufficient because baseline covariates may not always predict the final clinical outcome or principal stratum status. Methods estimating the potential outcome or principal stratum via both baseline covariates and post-baseline intermediate outcomes, in spite of some strong assumptions, seem more reasonable than the other three approaches.

We consider 3 principal strata (populations) that are of interest in general:

- $S_{**} = \{A(0) \in \{0,1\}, A(1) \in \{0,1\}\}$, which is the population of all patients
- $S_{*+} = \{A(1) = 1\}$, which is the principal stratum for patients that are adherent to treatment $T = 1$, $regardless\ of\ adherence\ to\ T = 0$

- $S_{++} = \{A(0) = 1, A(1) = 1\}$, which is the principal stratum for patients that are adherent to both treatments

The population $S_{**}$ is not of interest in this tripartite approach but, for comparison, the treatment effect will be estimated for this population in the example in Section 3. The choices of $S_{*+}$ and $S_{++}$ are based on the arguments presented by Qu et al [6] and Permutt [10]. For placebo-controlled studies, $S_{*+}$ is clinically meaningful as trial stakeholders are more interested in the population of patients who can adhere to an investigational treatment, regardless of adherence to placebo. For active controlled trials, the target population for the estimand of interest is $S_{++}$. In this situation, one can imagine that a physician is making a choice between treatments: the selection of which treatment will produce a more efficacious outcome should be based on the principal stratum of patients who could adhere to both treatments. Of course, if it is known that one treatment also increases the likelihood of adherence, such trade-offs need to be considered when making prescribing decisions.

In this article, we applied ACEs for the treatment difference for those who can adhere to one treatment or both treatments based on a causal inference framework [6]. These estimators are consistent for the treatment effect in the adherent strata of interest under some assumptions, and simulations demonstrated that the ACEs provide consistent estimates of the treatment difference for sample sizes that are typical for clinical trials. As the methods are complex and have been previously described [6], here we only provide a high-level description of the methods. The following assumptions have been posed for the validity of the ACEs:

A1: $Y = Y(1)T + Y(0)(1 - T)$

A2: $Z = Z(1)T + Z(0)(1 - T)$

A3: $A = A(1)T + A(0)(1 - T)$

A4: $T \perp \{Y(1), A(1), Z(1), Y(0), A(0), Z(0)\}|X$

A5: $A\{i\} \perp \{Y(1), Y(0), Z(1 - i)\}|\{X, Z(i)\}, \quad \forall i = 0,1$

A6: $Y(i) \perp Z(1 - i)|\{X, Z(i)\}, \quad \forall i = 0,1$

A7: $Z(0) \perp Z(1)|X$

The sign "⊥" denotes "statistically independent". Assumptions A1-A3 are the *stable unit treatment value assumptions*, i.e., the hypothetical outcome is the same as the observed outcome under the same treatment, which are reasonable and standard in causal inference approaches. Assumption A4 is the *ignorable treatment assignment assumption* [12]. Assumption A5 is the *ignorable adherence assumption*, which means adherence only depends on observed values. This means that the probability of ICEs can be modelled through the observed data (e.g., occurrence of an AE or deterioration of efficacy) during the clinical trial. We believe this is reasonable since patients are likely to continue their study treatment for the duration of a trial unless they are experiencing some unsatisfactory response, which we can observe or deduce from observed data, or for some administrative reason completely unrelated to treatment or baseline characteristics. Assumption A6 means, conditional on the baseline values and the intermediate outcome under one treatment, the outcome for this treatment is independent of the potential intermediate outcome in the other treatment. Assumption A7 assumes the potential intermediate outcomes under the 2 treatments are independent given baseline covariates. The dependencies between treatment, the baseline covariate, the intermediate outcome, the primary outcome, and the adherence are illustrated by the causal diagram in Figure 1.

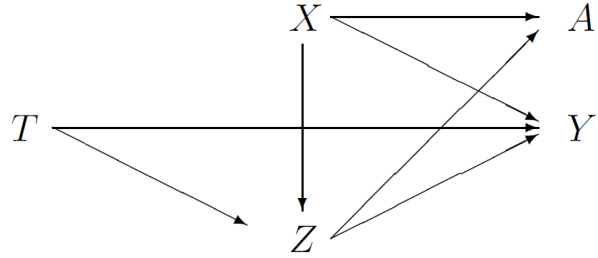

Figure 1: Causal diagram showing the dependencies between treatment ($T$), baseline covariate ($X$), post-baseline intermediate variable ($Z$), adherence ($A$), and outcome ($Y$).

Under Assumptions A1-A7, a consistent estimator for the treatment difference for stratum $S_{*+}$ is given by

$$\frac{1}{n_{11}} \sum_{j \in \{T_j=1, A_j=1\}} Y_j - \frac{1}{n_{11}} \sum_{j \in \{T_j=1, A_j=1\}} \hat{\phi}_0(X_j), \tag{1}$$

where $j$ is the subject index, $n_{11}$ is the number of patients who are adherent to the treatment when they take the experimental treatment, and $\hat{\phi}_t(X_j)$ is the estimated potential response for subject $j$ if taking treatment $T = t, (t = 0,1)$. The function $\phi_0$ is estimated using the data in the reference treatment group by first building a regression model of $Y$ on $X$ and $Z$, and then conditioning on $X$. Then, the potential outcome under the reference treatment for subjects assigned to the experimental treatment, also called "virtual twin", is estimated and denoted as $\hat{\phi}_0(X_j)$. Finally, the treatment difference for $S_{*+}$ can be estimated by taking the average of the difference between the observed outcome for the adherent patients on the experimental treatment and the potential outcome under the experimental treatment for those same patients in the reference treatment group.

A consistent estimator for the treatment difference for stratum $S_{++}$ is given by

$$\frac{\sum_{j\in\{T_j=0,A_j=1\}} \hat{\varphi}_1(X_j)}{\sum_{j\in\{T_j=0,A_j=1\}} \hat{h}_1(X_j)} - \frac{\sum_{j\in\{T_j=1,A_j=1\}} \hat{\varphi}_0(X_j)}{\sum_{j\in\{T_j=1,A_j=1\}} \hat{h}_0(X_j)}, \quad (2)$$

where $\hat{\varphi}_t(X_j)$ is the expected value of the product of the probability of adherence and the response of $Y$, conditional on $X_j$ if patient $j$ takes treatment $T = t$, and $\hat{h}_t(X_j)$ is the expected probability of adherence, conditional on $X_j$ if patient $j$ takes treatment $T = t$. For the ACE estimator in (2), in addition to the assumption for the ACE estimator in (1), we also assume the principal score (i.e., the probability of treatment adherence) is a function of baseline covariates ($X$) and intermediate outcome ($Z$). The idea is that the estimator for the mean response for treatment $T = t$ on $S_{++}$ is calculated using the weighted average of the potential outcome on the alternative treatment (i.e., $T = 1 - t$) by the probability of adherence to the alternative treatment ($T = 1 - t$). The probability of adherence to the alternative treatment can be estimated using the baseline covariates and the potential intermediate outcome under the alternative treatment, which is estimated in a similar fashion as $\hat{\phi}_t(X_j)$. The $\hat{\varphi}_t(X_j)$ is the expected value of the product of this probability of adherence to the alternative treatment and $\hat{\phi}_t(X_j)$, conditional on $X$. Again, the details for the construction of $\hat{\phi}_t, \hat{\varphi}_t$, and $\hat{h}_t$ have previously been described in detail [6].

Note the principal stratum of patients who would adhere to both treatments ($S_{++}$) cannot be observed. One estimator for the proportion of patients that would adhere to both treatments ($p_{++}$) is

$$\hat{p}_{++} = \frac{\sum_{j\in\{T_j=1,A_j=1\}} \hat{h}_0(X_j) + \sum_{j\in\{T_j=0,A_j=1\}} \hat{h}_1(X_j)}{n_1 + n_0}. \quad (3)$$

## 3. Application

### 3.1. The study and data

The data used in this article are based on the IMAGINE-3 Study: a 52-week, multi-center, phase 3 study of patients with type 1 diabetes mellitus. This was a parallel, double-blind study with randomization of qualified patients to basal insulin lispro (BIL) versus insulin glargine (GL). In this trial, 1114 adults with type 1 diabetes were randomized to BIL and GL in a 3:2 ratio (664 in BIL: 450 in GL), stratified by baseline hemoglobin A1c (HbA1c) (≤8.5%, >8.5%), baseline low-density lipoprotein cholesterol (LDL-C) (<100 mg/dL [2.6 mmol/L], ≥100 mg/dL), and prior basal insulin therapy (GL/insulin detemir/other basal insulin). In the first 12 weeks of the titration period, BIL or GL dose was adjusted weekly. Patients then entered the maintenance period (Weeks 12 to 52). During the study, patients were not allowed to take additional anti-diabetes medication unless they discontinued the randomized study treatment. Therefore, we consider treatment discontinuation as the only ICE for this study. The stated primary objective of the clinical trial was to demonstrate superiority of BIL to GL on HbA1c after 52 weeks of treatment. The study was conducted in accordance with the International Conference on Harmonization Guidelines for Good Clinical Practice and the Declaration of Helsinki. All patients signed an informed consent document, and the protocols and consent documents were approved by local ethical review boards prior to study initiation. This study was registered at clinicaltrials.gov as NCT01454284 and details of the study report have been published [24].

This post hoc analysis was conducted on all randomized patients who took at least one dose of the study drug (n=1112). One patient in each treatment arm did not take any study drug after randomization. Per the study design and data collection instruments, patient disposition was classified into 7 categories: AE, death, lost to follow-up, protocol violation, withdrawal by subject, physician decision, or sponsor decision. Detailed AE reports and some free-text comments from investigators suggested that the above 7 categories did not always reflect the real reason for discontinuation. For example, some patients who were discontinued due to LoE based on written comments in the case report form and laboratory measures of blood glucose were recorded as "physician decision." Furthermore, some patients who were discontinued due to a protocol specified AE were recorded as "sponsor decision." Therefore, using all data available, we re-assessed the treatment discontinuations to define the variables ($I_{AE}$ and $I_{LoE}$) related to the first 2 estimands. Based on the definitions of $I_{AE}$ and $I_{LoE}$, it is possible that a patient may have 2 simultaneous ICEs: one due to AE and one due to LoE.

In order to accurately define $I_{AE}$ and $I_{LoE}$, we manually reviewed data for each individual patient who discontinued the study or permanently discontinued the study treatment. The data included investigator provided reasons for discontinuation, investigators' comments on the discontinuation, efficacy data (HbA1c and glucose at baseline and prior to discontinuation), and safety data including adverse events and laboratory data (e.g., lipid panel and liver enzymes collected within 90 days before discontinuation). A conservative approach was implemented to classify early discontinued patients into the ICE category of "due to AE" if there were any recorded data or comments suggesting such was the case. Specifically, a patient's discontinuation was classified as an ICE due to AE if the investigator indicated it was "due to AE", or there were obvious safety issues such as the emergence of adverse events and/or abnormal laboratory data which could lead to discontinuation. A patient's discontinuation was classified as an ICE due to LoE if his or her HbA1c or glucose values at discontinuation were believed to have had no meaningful improvement from baseline values. A patient's discontinuation was classified as an ICE due to administration if there was no obvious evidence showing the discontinuation was due to an AE or LoE. Figure 2 summarizes the ICE classification criteria.

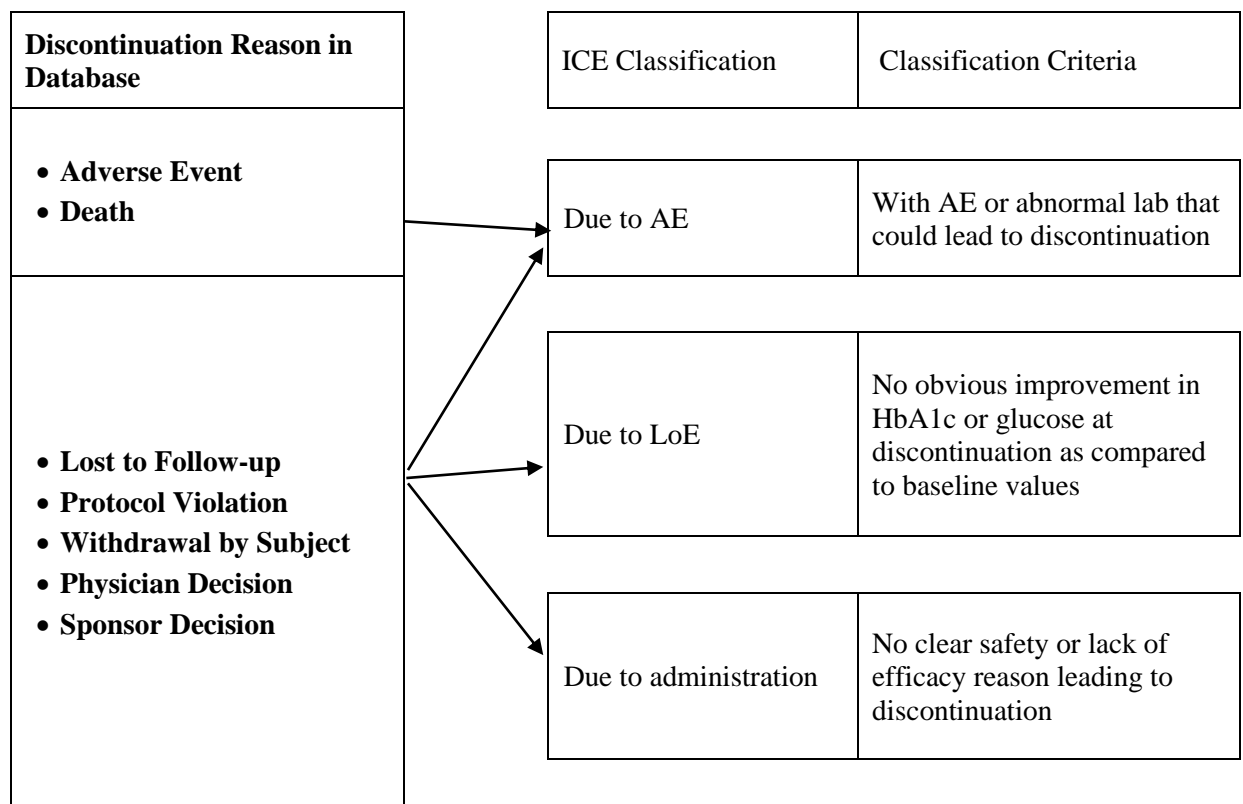

Figure 2. ICE classification criteria

3.2. Proportion of patients who had the first ICE due to AE and LoE (Estimands 1 and 2)

Among the 1114 randomized patients, 1112 patients (663 in BIL, 449 in GL) took at least one dose of study drug. Baseline characteristics for all randomized patients were balanced between the two treatment groups as expected [24]. During the 52 weeks of treatment, 235 patients experienced at least one ICE before completing the 52-week treatment period: 94 (8.5%) patients had the first ICE due to AE, 29 (2.6%) patients had the first ICE due to LoE, 120 (10.8%) patients had the first ICE due to administrative reasons, and 877 (78.9%) patients adhered to the treatment (Table 1). Significantly more patients had the first ICE due to AE in the BIL treatment group compared to GL (10.6% vs 5.3%; p=0.002). The disparity of ICEs due to AE was primarily due to injection site reactions [24]. The percentages of patients who had the first ICE due to LoE were similar between BIL and GL (2.7% vs 2.4%; p=0.850). The percentages of patients who had the first ICE due to administrative reasons were also quite similar between treatment groups (10.6% vs 11.1% for BIL vs GL, respectively; p=0.768). The cumulative incidences of the first ICE due to each reason over time are presented in Figures 3(a), 3(b), and 3(c).

For ICEs due to LoE, an important question from patients is "How long do I have to wait to know if I will respond or not to treatment?" Figure 3(d) summarizes the proportion of the first ICE due to LoE by time interval for each treatment. Among those who had the first ICE due to LoE, approximately half occurred during the first 3 months, approximately two thirds during the first 6 months, and the remaining third occurred over the last 6 months of the trial.

Table 1. Summary and Comparison of Proportion of Patients with Categories of First ICEs

| Disposition | GL (N=449) n (%) | BIL (N=663) n (%) | Difference (%, 95% CI) (BIL vs GL) | P-value |
|---|---|---|---|---|
| All Patients with ICEs, n (%)[†] | 81 (18.0) | 154 (23.2) | 5.2 (0.4, 10.0) | 0.043 |
| First ICE due to AEs, n (%) | 24 (5.3) | 70 (10.6) | 5.2 (2.1, 8.3) | 0.002 |
| First ICE due to LoE, n (%) | 11 (2.4) | 18 (2.7) | 0.3 (-1.6, 2.2) | 0.850 |
| First ICE due to Administrative Reasons, n (%) | 50 (11.1) | 70 (10.6) | -0.6 (-4.2, 3.2) | 0.768 |

Abbreviations: AE, adverse event; BIL, basal insulin lispro; CI, confidence interval; GL, insulin glargine; ICE, intercurrent event; LoE, lack of efficacy.
[†]Eight patients had concurrent first ICE due to AE and ICE due to LoE

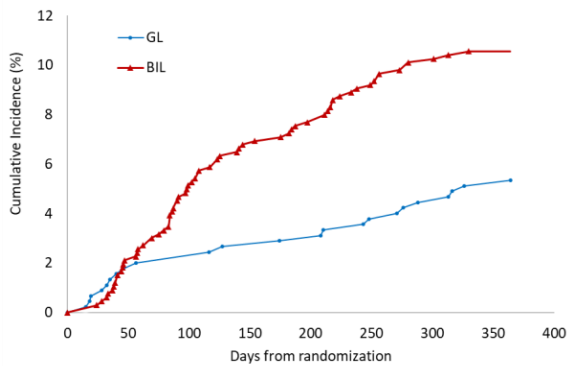

(a) Cumulative Incidence of the first ICE due to AE

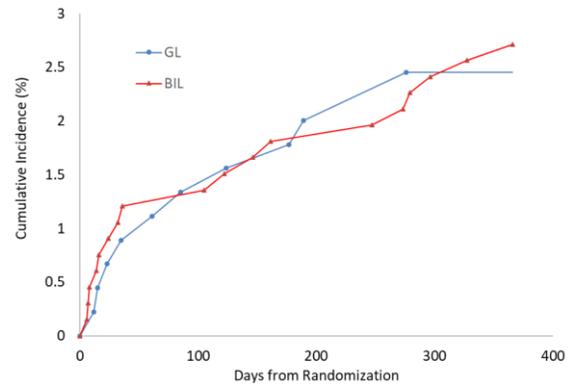

(b) Cumulative incidence to the first ICE due to LoE

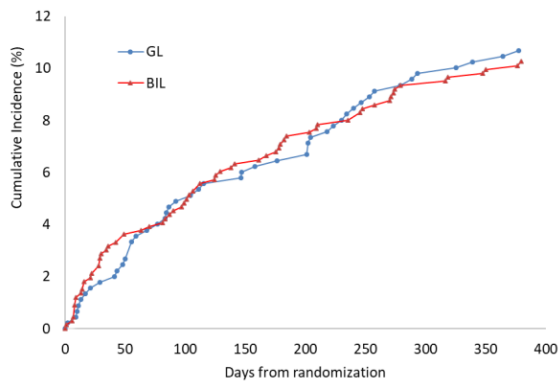

(c) Cumulative incidence for the first ICE due to administration

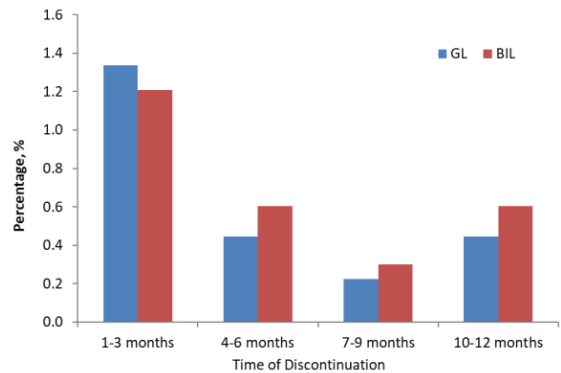

(d) The percentage of patients with the first ICE due to LoE by time period

Figure 3. Cumulative incidence of the first ICE (panels (a), (b), and (c)) and a histogram of time of the first ICE due to LoE (panel (d)).

3.3. Baseline characteristics for adherers and non-adherers

There are two other important comparisons of baseline characteristics in this analysis using the tripartite estimand approach, especially related to Estimand 3:

1) A comparison of those patients who discontinued the study treatment with those patients who adhered to the study treatment in order to understand if the discontinued patients might be different from the adherers. This can provide insight into how the adherent stratum might differ from the all randomized patient population;
2) For those who adhered to the assigned treatment for the duration of the study, the baseline characteristics for adherers between BIL and GL were compared to explore potential differences between the 2 adherers sets. While this gives insight into the comparability of the study treatments for the *observed* group of adherers, it is recognized that there may be unmeasured baseline covariates that are imbalanced since these patients are no longer randomized groups.

Table A.1 in the Appendix summarizes the demographic and baseline characteristics for the 4 groups (3 categories of first ICEs and adherers). Statistical comparisons were conducted between adherers and non-adherers. Compared to adherers, non-adherers were slightly more likely to be female, younger with lower body weight, have higher baseline HbA1c and triglyceride values, and lower total bilirubin. Furthermore, the non-adherers were more likely to be current smokers and to use statins, and less likely to be from Europe. The predictability of adherence using these baseline variables with multiplicity adjustment was also evaluated using a gradient boosting model [20]. Results showed the most influential baseline covariate on adherence was the age of the patient with a relative importance of 31% (adjusted p-value = 0.182) (Figure A.1). Thus, there were some differences between the adherers and non-adherers, but the differences did not seem to be dramatic.

Table A.2 in the Appendix summarizes and compares the demographics and baseline characteristics of the adherers by their respective treatment groups. Demographic and baseline characteristics were very similar for patients who adhered to the study treatment between the BIL and GL groups, with the only possible exception being estimated glomerular filtration rate (eGFR) (p=0.09). This slight difference was viewed as having minimal, if any, clinical significance in assessing the HbA1c response. In addition, all these baseline characteristics were included in a subgroup identification analysis to evaluate potential baseline variables that could predict differential adherence status between treatment groups with multiplicity-adjusted p-values [25]. The results showed that none of these measures were predictive of who would discontinue study treatment across treatment groups (baseline eGFR had the smallest multiplicity adjusted p-value of 0.242). Thus, despite being self-selected groups of patients, those patients who adhered to their GL and BIL treatment assignment were very similar with respect to *measured* baseline characteristics.

## 3.4 Efficacy for adherers (Estimand 3)

There were 76.8% and 82.0% of patients who adhered to BIL and GL during the 52 weeks of treatment, respectively. As mentioned previously, direct comparison of the efficacy for the adherers between the 2 treatments was not recommended because it does not estimate any *causal* estimand.

To use the methods proposed by Qu et al to model the probability of adherence and predict the post-baseline outcomes, we consider the following baseline covariates ($X$) that could potentially impact treatment adherence: age, gender, HbA1c, LDL-C, triglyceride (TG), fasting serum glucose (FSG), and alanine aminotransferase (ALT). The study also collected HbA1c, LDL-C, TG, FSG, and ALT at Week 12 and Week 26, and injection site reaction adverse events (a binary variable) that occurred between randomization and Week 12 and between Week 12 and Week 26. Those 6 post-baseline variables were considered in intermediate covariates $Z_1$ for Week 12 and $Z_2$ for Week 26, respectively. The probability of adherence was estimated using a multiplicative probability model [6].

Estimates of the treatment effect for the two principal strata of interest ($S_{*+}$ and $S_{++}$) are provided in Table 2. In Table 2, we also present other commonly used methods:

- Naïve adherers analysis. This analysis makes a simple comparison of treatment effect on the adherers in one treatment group to the adherers in another group without considering the potential imbalance for the adherers between the two treatment groups. Using our previous notation, this estimator tries to estimate
$$E\{Y(1)|A(1) = 1\} - E\{Y(0)|A(0) = 1\},$$
which apparently does not have a clear clinical meaning since it conditions on 2 different patient strata.
- Mixed model with repeated measures (MMRM) analysis using all observed data prior to the occurrence of ICEs. This analysis estimates the treatment difference in HbA1c at 52 weeks for all randomized patients, assuming all patients *would* adhere to the treatment without ICEs, i.e.,

$$e_4 = E\{Y(1) - Y(0)|S_{**}\},$$

which is an estimand based on the hypothetical strategy. This analysis assumes the missing data (either unobserved or censored by ICEs) are random.

- MMRM analysis with jump to reference (J2R) multiple imputations [26]. In this analysis, which includes the observed data prior to the occurrence of ICEs, the missing data (either unobserved or censored by ICEs) in the experimental treatment group were first imputed using the observed data in the reference arm, and then the MMRM model is applied to the imputed dataset. This analysis also estimates the estimand $e_4$, but assumes missing data are not missing at random.

Table 2. Estimates for various estimands for HbA1c at 52 weeks and the treatment difference

| Method | GL (LS mean ± SE) | BIL (LS mean ± SE) | Treatment difference for BIL vs GL (95% confidence interval) |
|---|---|---|---|
| Naïve adherers estimator | 7.57 ± 0.04 | 7.34 ± 0.03 | -0.23 (-0.33, -0.14) |
| ACE on $S_{*+}$ | 7.59 ± 0.05 | 7.34 ± 0.04 | -0.25 (-0.35, -0.15) |
| ACE on $S_{++}$ | 7.55 ± 0.05 | 7.31 ± 0.05 | -0.24 (-0.37, -0.10) |
| MMRM on $S_{**}$ | 7.61 ± 0.04 | 7.38 ± 0.03 | -0.22 (-0.32, -0.12) |
| MMRM after J2R imputations on $S_{**}$ | 7.62 ± 0.04 | 7.41 ± 0.03 | -0.21 (-0.30, -0.11) |

Abbreviations: ACE, adherers causal estimator; BIL, basal insulin lispro; GL, insulin glargine; J2R, jump to reference; LS, least squares; MMRM, mixed model with repeated measures; SE, standard error.

All estimates had similar results with the difference between estimates not exceeding 0.04%. It is not surprising to see the estimate for the naïve adherer set was similar to the estimate for stratum $S_{++}$ and $S_{*+}$, as the baseline characteristics for the adherers between the GL and BIL treatment groups were similar. The estimates for stratum $S_{++}$ were fairly similar to the estimates for the MMRM analyses with and without imputations. However, the ACEs are estimators based on principal stratification and they are fundamentally different from other estimators. The similarity between various estimates for different estimands is not surprising, as the ICEs due to AE (mostly injection reactions) or administrative reasons were not considered to be related to the efficacy based on the mechanism of action of both insulin treatments.

## 3.5. Graphical display of the tripartite estimands results

With the analyses in Sections 3.2 and 3.4, the tripartite estimands can be displayed in Figure 4. In the GL treatment arm, 24 (5.3%) patients discontinued study medication due to AE with a mean treatment exposure of 23.1 weeks, and 11 (2.4%) patients discontinued study medication due to LoE with a mean treatment exposure of 16.5 weeks. In the BIL treatment arm, 70 (10.6%) patients discontinued the study medication due to AE with a mean treatment duration of 12.3 weeks, and 18 (2.7%) patients discontinued study medication due to LoE with a mean treatment exposure of 19.6 weeks. For those who adhered to the BIL treatment (i.e., $S_{*+}$), the treatment difference (BIL vs. GL) was -0.25% (95% CI: -0.35%, -0.15%). The expected proportion of patients that would adhere to both treatments ($S_{++}$) was estimated to 61.9% and the treatment difference for these (potential) patients was -0.24% (95% CI: -0.37%, -0.10 %). Regardless of whether the tripartite approach is used in the analysis or any form of ACE is used for estimation, the graphic presentation like Figure 4 provides a useful summary of for treatment discontinuations in a clinical study.

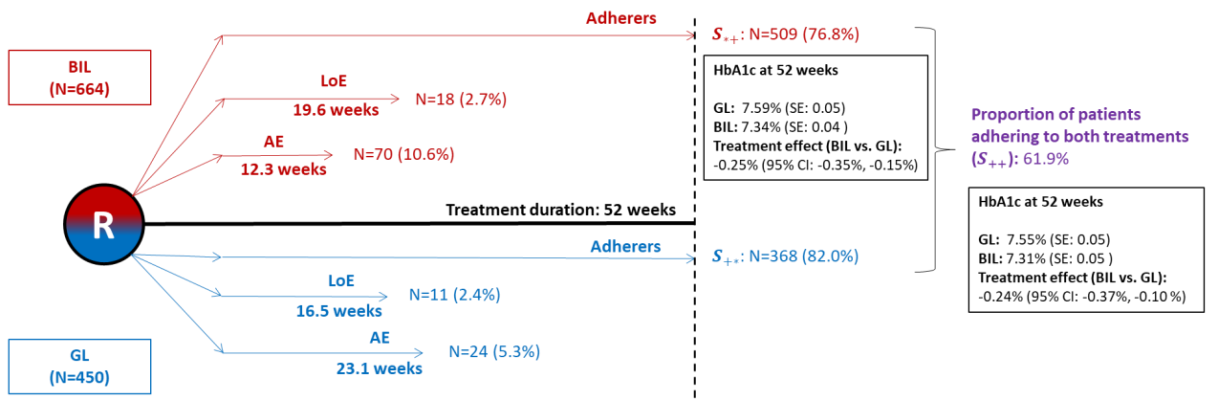

Figure 4. Visual display of the tripartite estimands. Abbreviations: AE, adverse events; BIL, basal insulin peglispro; CI, confidence interval; GL, basal insulin glargine; LoE, lack of efficacy; $S_{*+}$, a subset of patients those who adhered to the BIL treatment; $S_{++}$, a subset of patients those who would adhere to both treatments; SE, standard error.

## 4. Summary and Discussion

ICH E9 (R1) provides guidance on defining estimands, with a major focus on 5 specific strategies. All strategies except for the principal stratum approach focus on the treatment effect in all randomized patients. In regulatory interactions and key medical publications, there are almost no examples of using the principal stratum estimands for the primary analysis, with intent-to-treat (ITT) being the default strategy, enhanced by its recommendation in ICH-E9. While estimation for all randomized patients has been used for decision-making, we believe there are many situations when the estimands for principal stratum, especially based on the treatment adherence status, may be of interest or may even be considered for the primary estimands.

The most important qualifier for any estimand is that it should be first and foremost clinically meaningful. If the statistical design of clinical trials and the ensuing analysis is not aligned with what is clinically

meaningful – both to prescribers and patients – then even the finest statistical science is at best misguided and at worst self-serving. Undoubtedly, the interplay between what is clinically meaningful and what is statistically estimable requires thoughtful collaboration between clinicians, regulators, and statisticians [4, 27]. ICH E9 (R1) rightly states, "It should be agreed that reliable estimation is possible before the choice of estimand is finalized." Of course, "reliable estimation" is a subjective term; however, paraphrasing John Tukey, an approximate answer to the right question is much more desirable than a precise answer to the wrong question [28]. As ICEs and missing values exist, some assumptions, which often cannot be verified, are almost always required to derive consistent estimators for the estimand.

In some sense, an estimand based on the all randomized patient population is an unconditional estimand; that is, a description of what happens on average to everyone, a mixture of patients who adhere to their randomized treatment and those who do not adhere. It answers the question, "What is the expected treatment effect before the patient starts the treatment?" or "What is the expected treatment effect based on initiation of treatment?". In the tripartite approach, Estimand 3 is a conditional estimand which provides an answer to an important question, "What is the expected treatment effect in patients who can adhere to their treatment?" The formulation of the paper is related to the estimand in the principal stratum of patients who can adhere to treatment for the duration of the study. We suspect that most patients initially intend to adhere to a newly prescribed medication and want to know the expected outcome under that condition, even if they know or suspect there is a chance that they cannot adhere for one reason or another.

Of course, at the initiation of a new treatment, no one knows how an individual patient will respond. The question is, what should a physician communicate to the patient, or what should the sponsor/regulator communicate to the physician, about the expected response to treatment. We perceive the estimates provided by the tripartite estimands, when considered collectively, are more nuanced and informative for making benefit-risk decisions about initiating a new treatment. Furthermore, if they are indeed a more complete description of the treatment effect for the primary customer and consumer of the treatment, then perhaps they should also play a central role in regulatory decision-making and labeling.

A treatment can have an effect on those patients who take it, but also has an effect on which patients can take it in the first place. A complete description of the treatment effect includes both. While ICH E9 (R1) mostly discusses the role of ICEs in defining estimands for the primary outcome, it also mentions the value of defining estimands for ICEs themselves (see for example Section A.3.4). The tripartite estimands define 2 estimands for ICEs: the treatment differences in the proportion of patients with the first ICE due to AE and the proportion of patients with the first ICE due to LoE. This provides a more holistic approach in assessing efficacy and safety for decision-making by all stakeholders. This exercise also illuminated us on the ambiguities and potential deficiencies in capturing the reasons for discontinuation of study medication. We believe that the industry would benefit from redesign of standard case report forms for treatment discontinuation that are more explicit in capturing the reasons for discontinuation study medication, especially if it is for adverse events or lack of efficacy.

We applied the causal tripartite estimands retrospectively to a phase 3 clinical trial. For Estimand 3, the treatment difference for adherers, we used the recently developed ACEs with assumptions that we believe to be more reasonable than the monotonicity assumption used by most causal-inference estimation for adherers. We understand the assumptions for ACEs are still relatively strong, especially for Assumptions A6 and A7. Assumptions A6 and A7 are used to estimate the principal stratum. In commonly used sensitivity analyses for imputing the unobserved outcome, some sensitivity parameters (often conservative) are used to understand the robustness of the results; however, for estimating the principal stratum, it is difficult to say what is "conservative". Further research is required to better understand this assumption and some related sensitivity analyses. We expect the research in this area will continue to evolve, and more robust estimators may be available in the future.

In the example presented in this article, we found there were some differences in the baseline characteristics between adherers and non-adherers, but there was little difference in baseline characteristics between treatments for the observed adherers. In addition, the estimates for the treatment difference for the principal stratum for patients who could adhere to both treatments did not seem very different from the estimates for the estimand for all randomized patients based on hypothetical strategies. This is not surprising for several reasons.

Firstly, both treatments were the same class of drugs (daily basal insulin) and the expected treatment difference in the primary outcome of HbA1c was small (noting the primary objective of the IMAGINE-3 study was for non-inferiority). BIL was the first insulin showing greater reduction in HbA1c compared to another basal insulin, although the difference was only 0.2 to 0.3%. In addition, the lack of difference in the proportions of the first ICE due to LoE also suggested the difference in efficacy would likely to be small.

Secondly, the majority of AEs were not related to the study medication and were probably not related to efficacy measurements, as in most clinical trials.

Thirdly, the difference in the proportion of patients with the first ICE due to AE was primarily due to injection site reaction, which was treatment dependent, but was considered unlikely to be related to efficacy from a medical perspective.

Fourthly, the majority of patients were adherent to the assigned treatment, so the difference between the adherers' estimand and an estimand based on the hypothetical strategies was expected to be small.

In summary, lack of dramatic differences between different estimates, between treatments for baseline characteristics for adherers, and between adherers and non-adherers were not a sign of failure of ACEs, although failing on some assumptions might also result in indifferent outcomes. Indeed, unless there is a dramatic difference in ICEs that are correlated with efficacy, we expect similar "lack of difference" results will be present for the majority of clinical trials. In the simulation for the evaluation of the performance of ACEs [6], the simulation setting mimics the HbA1c change in real placebo-compared clinical trials. The probability of ICEs was modelled by baseline covariates and post-baseline intermediate efficacy measurements. Although there were differences in the proportion of patients with ICEs between treatments, the differences between estimates for the treatment difference for $S_{**}$, $S_{*+}$, and $S_{++}$ were less than 0.2%.

In conclusion, the trade-offs between the degree of efficacy conferred when taking a medication as prescribed and the occurrence of ICEs due to AE and LoE are essential and the tripartite estimands provide an opportunity for a more coherent evaluation of efficacy and safety, compared to the single estimand approach most often based on ITT currently in practice. ACEs, with the most reasonable assumptions among all existing estimators for estimands for principal stratum, can be used to estimate Estimand 3 of the treatment difference for adherers. Challenges remain, however, for the estimation of the estimand for adherers, including the assumptions and sensitivity analyses.

**Acknowledgement:** We would like to thank Dr. Ilya Lipkovich for his useful comments and suggestions for this manuscript.

# Appendix

Table A.1. Baseline characteristics by reason for disposition status

| Variable | DC due to AE N=94 | DC due to LoE N=128 | DC due to Admin N=120 | DC due to Any Reason N=235[†] | Adherers N=877 | P-value[‡] |
|---|---|---|---|---|---|---|
| Age | 40.55±13.55 | 41.83±13.78 | 38.80±14.48 | 39.70±13.90 | 42.48±13.15 | 0.004 |
| BMI (kg/m$^2$) | 25.92±3.51 | 26.85±3.70 | 26.34±4.54 | 26.23±4.08 | 26.56±3.91 | 0.247 |
| Body weight (kg) | 74.75±14.31 | 82.00±15.10 | 76.66±14.48 | 76.60±14.73 | 79.99±14.94 | 0.002 |
| Diabetes duration (years) | 19.43±13.00 | 25.98±17.19 | 17.36±12.01 | 18.83±12.72 | 20.01±12.49 | 0.201 |
| Baseline HbA1c (%) | 7.84±1.03 | 7.75±1.36 | 8.17±1.24 | 8.01±1.19 | 7.82±1.12 | 0.028 |
| HbA1c (%) right before DC | 7.50±1.09 | 8.08±1.36 | 7.75±1.15 | 7.68±1.17 | 7.43±1.01 | 0.003 |
| LDL-C (mg/dL) | 97.84±27.30 | 98.51±32.99 | 101.17±34.36 | 100.22±31.71 | 99.49±29.58 | 0.739 |
| Triglycerides (mg/dL) | 94.40±75.54 | 93.89±54.37 | 99.77±85.09 | 96.71±78.74 | 85.17±59.17 | 0.037 |
| Total cholesterol (mg/dL) | 182.34±32.46 | 182.95±36.28 | 183.19±38.07 | 183.44±35.98 | 179.33±34.25 | 0.106 |
| HDL-C (mg/dL) | 65.89±15.94 | 65.67±16.12 | 62.50±17.32 | 64.21±16.80 | 62.94±17.26 | 0.313 |
| FSG (mg/dL) | 176.41±77.88 | 170.21±66.15 | 170.49±82.95 | 172.86±79.80 | 171.73±78.02 | 0.844 |
| ALT (U/L) | 21.74±16.12 | 23.07±11.97 | 21.69±13.25 | 21.99±14.44 | 21.98±10.91 | 0.993 |
| AST (U/L) | 23.43±11.35 | 25.38±13.13 | 23.12±10.64 | 23.60±11.31 | 22.99±9.76 | 0.452 |
| Total bilirubin (umol/L) | 8.78±4.97 | 9.48±5.33 | 8.73±4.84 | 8.92±4.99 | 9.91±5.23 | 0.009 |
| Male | 41 (43.6) | 17 (58.6) | 64 (53.3) | 119 (50.6) | 559 (63.7) | <.001 |
| Region | | | | | | 0.018 |
|     European Union | 26 (27.7) | 2 (6.9) | 31 (25.8) | 59 (25.1) | 302 (34.4) | |
|     North America | 51 (54.3) | 22 (75.9) | 76 (63.3) | 143 (60.9) | 479 (54.6) | |
|     Other | 17 (18.1) | 5 (17.2) | 13 (10.8) | 33 (14.0) | 96 (10.9) | |
| Hypertension | 31 (33.0) | 10 (34.5) | 42 (35.0) | 79 (33.6) | 325 (37.1) | 0.36 |
| Lipid lowering medication | 29 (30.9) | 8 (27.6) | 39 (32.5) | 72 (30.6) | 319 (36.4) | 0.107 |
| Statin | 26 (27.7) | 7 (24.1) | 30 (25.0) | 60 (25.5) | 296 (33.8) | 0.018 |
| Smoking | | | | | | 0.006 |
|     Current user | 24 (25.5) | 5 (17.2) | 25 (20.8) | 53 (22.6) | 127 (14.5) | |
|     Never used | 58 (61.7) | 19 (65.5) | 74 (61.7) | 147 (62.6) | 570 (65.0) | |
|     Past user | 12 (12.8) | 5 (17.2) | 21 (17.5) | 35 (14.9) | 180 (20.5) | |
| eGFR | | | | | | 0.389 |
| ≥ 30 to < 60 | 8 (8.5) | 1 (3.4) | 9 (7.5) | 17 (7.2) | 68 (7.8) | |
| ≥ 60 to < 90 | 50 (53.2) | 16 (55.2) | 49 (40.8) | 111 (47.2) | 454 (51.8) | |
| ≥ 90 | 36 (38.3) | 12 (41.4) | 62 (51.7) | 107 (45.5) | 355 (40.5) | |
| Prior MI/CR/CABG | 4 (4.3) | 2 (6.9) | 0 (0.0) | 4 (1.7) | 20 (2.3) | 0.801 |

Mean±SD for continuous variables and the number (%) for categorical variables. Abbreviations: Admin, administration; AE, adverse event; ALT, alanine aminotransferase; AST, aspartate aminotransferase; BMI, body mass index; CABG, coronary artery bypass grafting; CR, cardiac rehabilitation; DC, discontinuation; eGFR, estimated glomerular filtration rate; FSG, fasting serum glucose; HbA1c, hemoglobin A1c; HDL-C, high-density lipoprotein cholesterol; LDL-C, low-density lipoprotein cholesterol; LoE, lack of efficacy; MI, myocardial infarction.

†Several patients discontinued due to the reasons of both of AE and LoE.

‡The comparison of adherers versus patient who discontinued their study medication using p-values is meant to be a descriptive in measuring how disparate or similar the two groups of patients are. These p-values should not be interpreted as leading to inferential conclusion relating to multiple hypothesis tests.

Table A.2. Baseline characteristics for adherers by treatment

| Variable | GL (N=368) | BIL (N=509) |
|---|---|---|
| Age | 43.10±12.84 | 42.04±13.36 |
| BMI (kg/m$^2$) | 26.73±4.00 | 26.45±3.85 |
| Body weight (kg) | 80.31±15.50 | 79.75±14.52 |
| Diabetes duration (years) | 20.64±12.65 | 19.55±12.36 |
| Baseline HbA1c (%) | 7.79±1.07 | 7.85±1.15 |
| HbA1c (%) right before DC | 7.55±1.04 | 7.34±0.98 |
| LDL-C (mg/dL) | 99.92±29.32 | 99.17±29.79 |
| Triglycerides (mg/dL) | 82.60±45.17 | 87.02±67.48 |
| Total cholesterol (mg/dL) | 179.48±33.59 | 179.23±34.76 |
| HDL-C (mg/dL) | 63.03±16.52 | 62.87±17.79 |
| FSG (mg/dL) | 168.97±76.24 | 173.72±79.30 |
| ALT (U/L) | 21.48±9.29 | 22.34±11.94 |
| AST (U/L) | 23.01±9.07 | 22.98±10.24 |
| Total bilirubin (umol/L) | 9.96±5.27 | 9.87±5.20 |
| Male | 233 (63.3) | 326 (64.0) |
| Region | | |
|     European Union | 133 (36.1) | 169 (33.2) |
|     North America | 194 (52.7) | 285 (56.0) |
|     Other | 41 (11.1) | 55 (10.8) |
| Hypertension | 144 (39.1) | 181 (35.6) |
| Lipid lowering medication | 137 (37.2) | 182 (35.8) |
| Statin | 127 (34.5) | 169 (33.2) |
| Smoking | | |
|     Current user | 58 (15.8) | 69 (13.6) |
|     Never used | 235 (63.9) | 335 (65.8) |
|     Past user | 75 (20.4) | 105 (20.6) |
| eGFR | | |
|     ≥ 30 to < 60 | 22 (6.0) | 46 (9.0) |
|     ≥ 60 to < 90 | 204 (55.4) | 250 (49.1) |
|     ≥ 90 | 142 (38.6) | 213 (41.8) |
| Prior MI/CR/CABG | 9 (2.4) | 11 (2.2) |

Mean±SD for continuous variables and the number (%) for categorical variables. Abbreviations: ALT, alanine aminotransferase; AST, aspartate aminotransferase; BIL, basal insulin peglispro; BMI, body mass index; CABG, coronary artery bypass grafting; CR, cardiac rehabilitation; DC, discontinuation; eGFR, estimated glomerular filtration rate; FSG, fasting serum glucose; GL, basal insulin glargine; HbA1c, hemoglobin A1c; HDL-C, high-density lipoprotein cholesterol; LDL-C, low-density lipoprotein cholesterol; MI, myocardial infarction.

**Figure A.1**. The relative importance of baseline covariates and treatment on adherence based on gradient boosting machine model (p-value was calculated based on a randomized permutation test).
Abbreviations: ALT, alanine aminotransferase; AST, aspartate aminotransferase; BMI, body mass index; CABG, coronary artery bypass grafting; CR, cardiac rehabilitation; eGFR, estimated glomerular filtration rate; FSG, fasting serum glucose; HbA1c, hemoglobin A1c; HDL-C, high-density lipoprotein cholesterol; LDL-C, low-density lipoprotein cholesterol; MI, myocardial infarction.

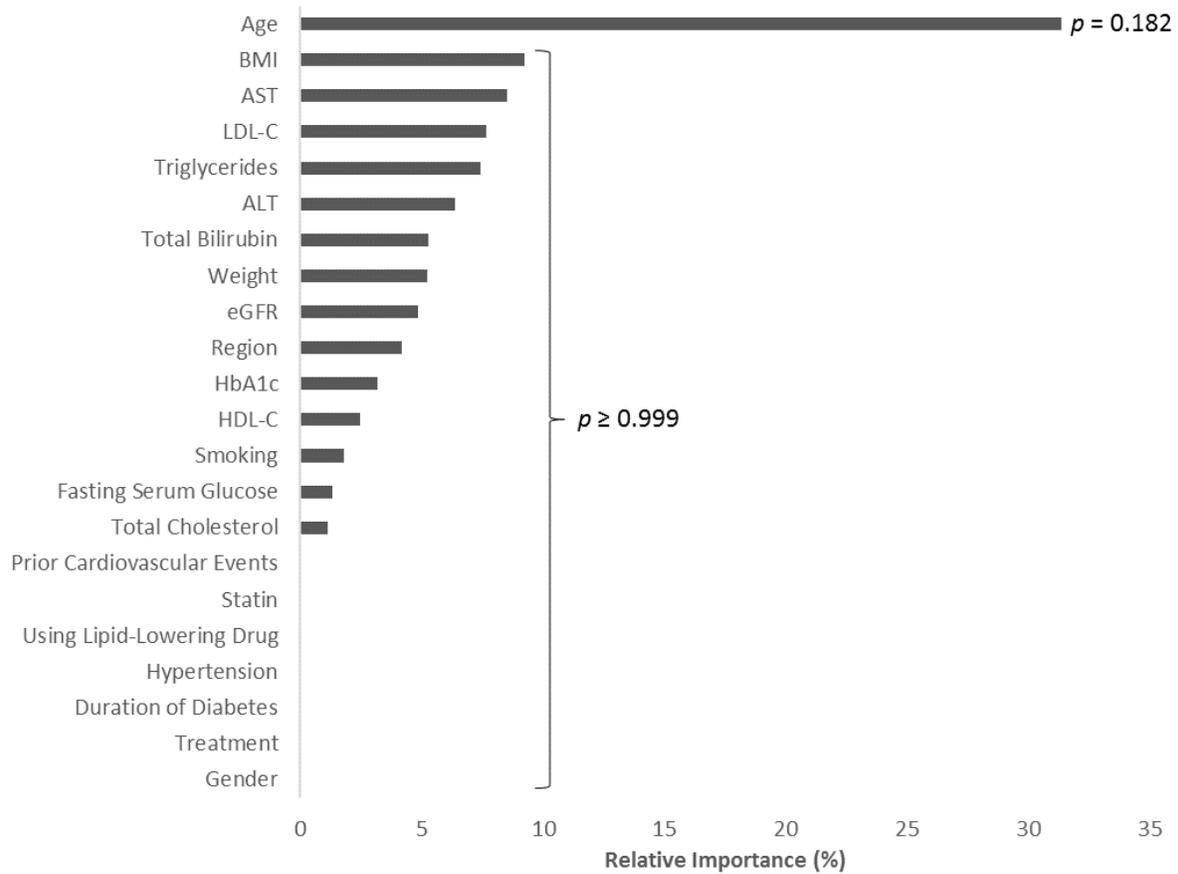